\documentclass[%
superscriptaddress,
showpacs,preprintnumbers,
nobibnotes,
 amsmath,amssymb,
 aps,
 pre,
 longbibliography,
 lengthcheck,%
]{revtex4-1}

\usepackage{graphicx}
\usepackage{dcolumn}
\usepackage{bm}
\usepackage{hyperref}

\begin{document}

\title{Normal-to-anomalous diffusion transition in disordered correlated potentials: from the central limit theorem to stable laws.}

\author{R. Salgado-Garc\'{\i}a} 
\email{raulsg@uaem.mx}
\affiliation{Facultad de Ciencias, Universidad Aut\'onoma del Estado de Morelos. Avenida Universidad 1001, Colonia Chamilpa, 62209, Cuernavaca Morelos, Mexico.
}

\author{Cesar Maldonado} 
\affiliation{Instituto de F\'{\i}sica, Universidad Aut\'onoma de San Luis Potos\'{\i}, Avenida Manuel Nava 6, Zona Universitaria, 
78290 San Luis Potos\'\i , Mexico.}

\date{\today} 

\begin{abstract} 
We study the diffusion of an ensemble of overdamped particles sliding over a tilted random potential (produced by the interaction of a particle with a random polymer) with long-range correlations. We found that the diffusion properties of such a system are closely related to the correlation function of the corresponding potential. We model the substrate as a symbolic trajectory of a shift space which enables us to obtain a general formula for the diffusion coefficient when normal diffusion occurs. 
The total time that the particle takes to travel through $n$ monomers can be seen as an ergodic sum to which we can apply the central limit theorem. The latter can be implemented if the correlations decay fast enough in order for the central limit theorem to be valid. On the other hand, we presume that when the central limit theorem breaks down the system give rise to anomalous diffusion. We give two examples exhibiting a transition from normal to anomalous diffusion due to this mechanism. We also give analytical expressions for the diffusion exponents in both cases by assuming convergence to a stable law. Finally we test our predictions by means of numerical simulations.

\end{abstract}


\maketitle

\section{Introduction}

During the last decade it has been an increasing interest in the transport properties of particles in disordered media. This has been mainly motivated by the study of biological transport like translocation of DNA  through a porous~\cite{branton2008potential}  as well as protein transport along DNA chains~\cite{slutsky2004diffusion,mirnyhow2009,gormanvisualizing2008}. Moreover, the theoretical basis on the understanding transport phenomena in disordered potentials could result in important technological applications such as particle separation at mesoscales, controlling colloidal particles in optical traps~\cite{reimann2008weak} or novel genome sequencing techniques~\cite{branton2008potential,ashkenasy2005recognizing}. 

The problem of how the particles move on disordered potentials has been tackled from several points of view. For example in Ref.~\cite{romero1998brownian} A. H. Romero and J. M. Sancho studied analytically and numerically the movement of Brownian motion of non-interacting overdamped particles on random potentials  with short-range correlations. Particularly, they found several regimes for diffusion which are explained in terms of the characteristics of the gaussian distribution of the potentials. In Ref.~\cite{kunz2003mechanical}, H. Kunz \emph{et al} introduced a simple mechanical model to understand how the quenched disorder gives rise to anomalous diffusion. They consider an ensemble of non-interacting overdamped particles that moves deterministically on a tilted random potential. This approach differs from previous works in the sense that thermal fluctuations are absent and the objective is to understand the origin of anomalous diffusion from the very deterministic dynamics. The randomness of the potentials in such a model was introduced as a series of disjoint random ``scatters'' put along a one-dimensional space where the motion occurs. A more recent approach to the problem of deterministic diffusion of overdamped particles in disordered potentials was carried by S. I. Denisov \emph{et al} in Ref.~\cite{denisov2010biased}. They consider a class of piece-wise linear random potentials and this simplification let them find analytical expression for the diffusion coefficient (when normal diffusion occurs) as well as for the long-time and the short time behavior of the mean square displacement. They are able in this way to characterize the corresponding anomalous diffusion. However, these deterministic approaches only consider uncorrelated potentials.  

On the other hand, recent works by G. Gottwald and  I. Melbourne~\cite{gottwald2012central,gottwald2013huygens} have shown the suppression of anomalous diffusion in group extensions of a kind of non-uniformly hyperbolic dynamical systems. This suppression is characterized by certain symmetries of the corresponding observable.  Actually the transition from normal to anomalous diffusion has often studied (rigorously) on dynamical systems having intermittent behavior. In this paper we will give an example of this phenomenon on a system which, up to our knowledge, is different in nature to the previously considered. Here we deal with the problem of deterministic diffusion of overdamped particles in a random potential with long-range correlations. Our model has significant differences with the previous approaches mentioned above. The system we consider here can be thought as a particle moving on a chain (e.g., a polymer). We assume that the chain is conformed of  ``cells'' or monomers of constant length. The monomers are assumed to be taken from a  finite (or countable infinite) set and the chain is built up by concatenating at random (by means of some stochastic process) copies of the possible monomers. To be more precise, the chain we have in mind can be a DNA sequence which consist of four types of monomers: adenine, cytosine, thymine and guanine.  We assume that the particle has a specific interaction with every monomer type, and this interaction defines the potential profile that feels the particle when is placed over the chain. Then, if we put an overdamped particle in such a chain and we apply an external force, the particle will move in a specific direction if the strength of the bias is large enough.  

This model of deterministic motion on a disordered potential lets us understand how normal and anomalous diffusion arise from the correlations. The main tool we use is the central limit theorem (CLT). This theorem, which is valid for ergodic sums of well-behaved observables in ergodic dynamical systems with sufficiently fast decay of correlations~\cite{chazottes2012fluctuations,gouezel2004sharp}, allows us characterize the asymptotic distribution of a sum of \emph{crossing times} of the particles through unit cells. This is later used to obtain the diffusion coefficient when the transport is normal. When the CLT is no longer valid for sufficiently slow decay of correlations it is expected to observe anomalous diffusion. In such a case a more general theorem is needed in order to obtain some expression for the asymptotic mean-squared displacement. 

The paper is organized as follows: In Sec.~\ref{sec:Model} we state explicitly the model we use for the correlated disordered potentials. In Sec.~\ref{sec:NormalDiffusion} we give sufficient conditions for the occurrence of normal diffusion and give a general formula for the diffusion coefficient. We test our prediction for the diffusion coefficient for particles on disordered polymers produced using a Markov chain and the expansion--modification (E-M) process. Sec.~\ref{sec:AnomalousDiffusion} is devoted to explore numerical and analytically some examples of the occurrence of anomalous diffusion when the conditions for normal diffusion are not fulfilled. In these cases we show that a transition from normal to anomalous diffusion occurs. Finally Sec.~\ref{sec:Conclusions} we give a brief discussion of our result and give the main conclusions of our work. Some appendices containing detailed calculations are included.

\section{The model}
\label{sec:Model}

\begin{figure}[t]
\begin{center}
\scalebox{0.35}{\includegraphics{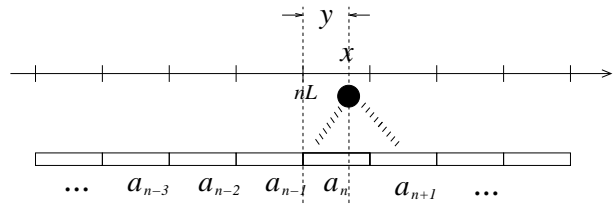}}
\end{center}
     \caption{
   Schematic representation of the particle-monomers interaction. When the particle is located at the $n$-th monomer the major contribution to the interaction comes from the $n$-th monomer and its neighbors. In the figure it is represented the case when the particle interacts with the $n$-th and the $(n+1)$-th monomers. In this case the interaction defines the potential profile that depends on the symbols $a_n$ and $a_{n+1}$ and the relative position $y$.  
     }
\label{fig:interaction_P-M}
\end{figure}

We will consider an overdamped particle moving on a one dimensional disordered  potential $V(x)$ subjected to an external force $F_0$. The equation of motion of such a particle is given by
\begin{equation}
\label{eq:overdamped_particle}
\gamma \frac{dx}{dt} = -V^\prime(x) + F_0,
\end{equation}
where $V(x)$ is the potential that feels the particle due to its interaction with the substrate where the motion occurs. The constant $\gamma $ is the friction coefficient. As we said in the introduction, the random media, or the substrate,  is assumed to be made up of unit cells of length $L$, called monomers hereafter. The possible types of monomers will be labeled with symbols from a set called $\mathcal{A}$. The cardinality of $\mathcal{A}$ is then the total number of different monomers and we assume that such a set can be finite or countable infinite. The substrate is then represented by a sequence of monomers as $\mathbf{a} := (\cdots, a_{-2}, a_{-1},a_0,a_1,a_2,\cdots)$ where $a_i\in \mathcal{A}$ is the $i$-th monomer. The monomers along the chain $\mathbf{a}$ are assumed to be produced by some stochastic process whose characteristics we will specify later. When the particle is placed at some position $x$, it feels a potential that results from its interaction with the monomers. Lets us  write the particle position as $x = nL + y$, where $n$ is the monomer at which the particle is located and $y$ is the relative position on the monomer (the particle position module $L$). Then, the interaction of the particle with the monomer is defined by specific kind of the $n$-th monomer, the closest monomer to the particle, and by its neighbors (see Fig.~\ref{fig:interaction_P-M}). In other words,  the potential  $V(x)$ is a function of $a_n$,  its closest  neighbors, and the relative position $y$.  For expository purposes, let us assume for the moment that the potential depends only on the $n$-th and on the $(n+1)$-th monomers and the relative position, i.e., $V(x) = \phi_{a_n,a_{n+1}} (y)$. 

%
%

It is easy to see that Eq.~\eqref{eq:overdamped_particle} can be solved to obtain the time $T_n$ the particle spend traveling a distance $nL$,  starting out at $ x = 0$,
\begin{equation}
T_n (a_0,a_1,\cdots, a_n)= \sum_{j=0}^{n-1} \tau (a_j,a_{j+1}) ,
\end{equation}
where $\tau(a,b)$ is the time that the particle takes to cross from the monomer $a$ to the monomer $b$ (called hereafter \emph{crossing time}), i.e., 
\[
\tau (a,b) = \gamma \int_{0}^L \frac{dy}{ -\phi^\prime_{a,b} (y)+F_0}.
\]  
Notice that, in order for $\tau$ be bounded it is necessary that the external forcing be such that $F_0 > -\inf \{\phi^\prime_{a,b}(x)\}$. We will assume that this is the case throughout  all this  work except in sub-section~\ref{subsec:Unbounded}, where we explore the occurrence of anomalous diffusion.

Let us now to specify the properties of the stochastic process which produce the substrate (the polymer). First assume that the chain $\mathbf{a}$ is infinitely long. In this case we can think of $\mathbf{a}$ as a point in a symbolic space  (a \emph{shift space}) which consists of all the infinite sequences made up with elements in the alphabet $\mathcal{A}$. This space is denoted by $\mathcal{A}^\mathbb{Z}$. If the infinite chain is randomly produced by certain stationary process, we need to specify a probability  measure $\mu$ (the stationary measure) on such a space.  We will assume that the  statistical properties of the random media are translationally invariant. Then, the probability measures that are compatible with this hypothesis are those which are invariant under the shift mapping. The shift mapping $\sigma : \mathcal{A}^\mathbb{Z} \to \mathcal{A}^\mathbb{Z}$ is defined as follows: if $\sigma (\mathbf{a}) = \mathbf{b}$ then $b_j = a_{j+1}$. An invariant  probability measure $\mu$ is such that for every $A$ in the Borel sigma algebra on $\mathcal{A}^\mathbb{Z}$, we have that~\footnote{For a review on symbolic dynamics see Ref.~\cite{lind1995introduction}.},
\[
\mu(A) = \mu( \sigma^{-1} (A)).
\]

The reason for modeling the substrate in this way is that we can consider more general situations than those considered in previous works. Indeed our method let us consider disordered chains produced in different ways. For example, we can built up symbolic chains by Markov stationary processes whose  corresponding measures are shift-invariant. Under the appropriate conditions these processes are ergodic and mixing~\cite{levin2009markov}. Another class of measures generating substrates that it is important to mention are the well-known Bowen-Gibbs measures. The latter are naturally shift-invariant~\cite{bowen2008equilibrium} and have been extensively studied within the thermodynamical formalism. In this case the substrate can be interpreted as a spin chain which is generated with a given Hamiltonian. The last example we consider that fits into our framework is the E-M system. This is one of the first models proposed to understand the statistical properties of DNA sequences~\cite{li1992long}. Recently it has been shown that the stochastic processes generating the symbolic sequences is actually a Markov process (in time) with a well-defined stationary measure that can be characterized exactly~\cite{salgado2012expansion}. 

\section{Normal Diffusion}
\label{sec:NormalDiffusion}

\subsection{The Central Limit Theorem}

In the theory of dynamical systems it is known that the CLT is valid for ergodic sums of regular observables with respect to the corresponding invariant measure. This is true if the correlations decay fast enough in such a way that the correlation function is absolutely summable~\cite{gouezel2004central,chernovlimit1995,chazottes2012fluctuations}. More precisely~\footnote{All the results stated here about the Central Limit Theorem can be found in Ref.~\cite{chazottes2012fluctuations}.}, in the context of shift spaces, given an observable $f: \mathcal{A}^\mathbb{Z} \to \mathbb{R}$ and given a \textit{typical} (with respect to $\mu$) symbolic sequence $\mathbf{a}$, the ergodic sum,
\[
S_n(\mathbf{a}) := \sum_{j=0}^{n-1} f(\sigma^j(\mathbf{a})),
\]
has typical fluctuations of order $\sqrt{n}$. Equivalently, we can say that given an $x \in \mathbb{R}$, the probability of all sequences $\mathbf{a}$ such that
\[
\frac{ S_n(\mathbf{a}) - n\int f d\mu }{\sqrt{ n}} \leq x ,
\]
is asymptotically given by the normal distribution, i.e., 
\[
\lim_{n\to \infty }\mu\left(  \frac{ S_n(\mathbf{a}) - n\int f d\mu }{\sqrt{ n}} \leq x  \right) =  \int_{-\infty}^x \frac{e^{-\frac{u^2}{2\varrho_f^2} }}{\sqrt{2\pi} \varrho_f }  du,
\]
where $\varrho_f^2$ is a non-zero constant representing the variance of the process $\{ S_n (\mathbf{a})\}_{n=0}^{\infty}$ divided by $n$ in the limit of $n\to \infty$.   Moreover, $ \varrho_f^2$ is a constant depending on the observable $f$, and it is not difficult to show that 
\begin{equation}
\label{eq:rho-correlations}
\varrho_f^2 = C_f(0) + 2 \sum_{\ell = 1}^\infty C_{f} (\ell).
\end{equation}
Here $C_f$ is the auto-correlation function of the observable $f$, i.e.,
\[
C_f(\ell) := \int f \cdot f\circ \sigma^\ell \mathrm{d}\mu - \left( \int f\mathrm{d}\mu\right)^2.
\]Finally we should stress that in order for $\varrho_f^2$ be given by Eq.~\eqref{eq:rho-correlations} it is sufficient that,
\begin{equation}
\label{eq:summable_corr}
\sum_{\ell = 1}^\infty \big | C_{f} (\ell) \big| < \infty,
\end{equation}
which is the summability condition mentioned above.

To prove the existence of normal diffusion in disordered potentials we will assume that the systems generating the substrates (the polymers) are ergodic and have summable correlations. In particular all the systems mentioned at the end of the last section satisfy both conditions. 

\subsection{The Diffusion Coefficient}

In our case, we identify the observable  with the crossing time $\tau$. Although the potentials were assumed to rely on two monomers, and consequently the function $\tau$ is a function depending on two coordinates of $\mathbf{a}$, we can consider the more general situation where the particles interacts with all the monomers in the chain. Under such a situation it is clear that the time to pass from one cell to adjacent one depends on all the ``labels'' in the symbolic sequence $\mathbf{a}$, i.e., $\tau : \mathcal{A}^\mathbb{Z} \to \mathbb{R}$.  

It is important to stress that the function $\tau (\mathbf{a})$ stands for the time to go from the  $0$-th monomer to the first one. If we want to calculate the time to go from the $j$-th monomer to the $(j+1)$-th monomer we need to ``shift'' forward the particle  up to the $j$-th cell and calculate the corresponding potential that feels  the particle there. This action is equivalent to shift backward the sequence  to the left, which is carried out by the shift mapping $\sigma$. Then, it is clear that the time that the particle spends traveling from the $j$-th monomer to the $(j+1)$-th monomer is given by $ \tau (\sigma^j(\mathbf{a}))$. Hence, given a symbolic sequence $\mathbf{a}\in \mathcal{A}^\mathbb{Z}$, the total time to go from the $0$-th monomer to the $n$-th monomer is given by,
\[
T_n (\mathbf{a}) = \sum_{j=0}^{n-1} \tau (\sigma^j(\mathbf{a})).
\]
Clearly, this quantity is an ergodic sum which under hypotheses mentioned above satisfies a CLT. To calculate the diffusion coefficient, which is defined by
\[
D = \lim_{t\to \infty} \frac{\mathrm{Var}[X_t]}{t},
\]
it is necessary to characterize the process $\{ X_t\}_{t\geq 0}$. The latter stands for the random trajectory of the particle. Here we will not characterize the particle position as a continuous process but instead as a discrete one. Indeed, our construction of the process lets us naturally  see the trajectory of the particle as a discrete process: the displacement of the particle by a distance $L$ (the monomer length) has associated a random time given by the function $\tau$. Thus, if we know the distribution of $T_n$ we can calculate the distribution of a random variable $N_t$ defined by the number of monomers which the particle has crossed during a time $t$. In this way we have that $X_t = LN_t $. The random variable $N_t$ is defined implicitly  by the equation $T_{N_t }= t$ for a positive $t$ fixed. By the CLT, it is clear that we can assume that $T_n$ has a normal distribution with mean $n\overline{\tau}$ and variance $n\varrho^2_\tau$ where the parameters $\overline{\tau}$ and $\varrho_\tau^2$ are given by,
\begin{eqnarray}
\overline{\tau} &:=& \mathbb{E}[\tau] = \int \tau \mbox{d} \mu,
\label{eq:mean_tau}
\\
\varrho^2_\tau &:=& \mbox{Var}[\tau] + 2\sum_{\ell = 1}^{\infty} C_\tau(\ell).
\label{eq:var_tau}
\end{eqnarray}
In this case, $C_\tau$ stands for the correlation function of the crossing times. In the Appendix~\ref{ape:Dist_N} it is shown that for large $t$, the distribution for $N_t$ is normal and indeed it can be written as,
\[
N_t = \frac{t}{\overline{\tau}} - \frac{\varrho_\tau \, t^{1/2} W}{\overline{\tau}^{3/2}},
\]
where $W$ is a random variable normally distributed with zero mean and variance equals one. The last equation lets us see that the mean displacement is given by $L\mathbb{E} [N_t] =  Lt/\overline{\tau}$, while the mean square displacement can be written as 
\[
 \mbox{Var}[X_t] = L^2 \mbox{Var}[N_t] = \frac{L^2\varrho^2_\tau \, t}{\overline{\tau}^3},
\]
for asymptotically large $t$. 
The latter implies normal diffusion and that the diffusion coefficient is given by
\begin{equation}
\label{eq:Deff}
D = \frac{L^2\varrho^2_\tau }{\overline{\tau}^3},
\end{equation}
whenever such quantity is finite.  In particular, if we have a polymer with long-range correlations, in order for the variance $\varrho_\tau^2$ to be finite it is necessary that the correlations decay sufficiently fast. Indeed, we need to meet the condition~\eqref{eq:summable_corr} for the observable $f = \tau$. This is true if the correlations decay faster than $\ell^{-\alpha}$ with $\alpha >1$.

\subsection{Examples}

Now we test our formula for the diffusion coefficient in some specific situations where the normal diffusion holds. 

\subsubsection{A Stationary Markov Chain}

Consider a set of three different monomers $\mathcal{A} = \{0,1,2\}$. Assume that an infinite polymer is built up at random by means of  a Markov process with stochastic matrix $P : \mathcal{A}\times \mathcal{A} \to [0,1]$ given by,
\[
P = \left(
  \begin{array}{ccc}
   0 & p & q \\
   q & 0 & p \\
   p & q & 0
  \end{array} \right).
\]
It is easy to see that this matrix is doubly stochastic and the unique invariant probability vector $\mathbf{\pi} = \mathbf{\pi} P $ is given by $\pi = (\frac{1}{3},\frac{1}{3},\frac{1}{3})$. We will model the interactions of a particle placed on such a polymer as follows. First we associate to each monomer a specific ``potential value''. The potential values corresponding to the monomers $0,1$ and $2$  are denoted by  $V_0,V_1$ and $V_2$. Then, the potential profile that feels the particle will be assumed to be piecewise linear and its corresponding slopes will be the potential differences between adjacent monomers divided by the monomer size. The gradient of the potential is then a piecewise constant function and it can then be seen as a matrix,
\[
F = \left(
  \begin{array}{ccc}
   0 & f_2 & f_3 \\
   -f_2 & 0 & f_1 \\
   -f_3 & -f_1 & 0
  \end{array} \right),
\]
where $f_1 := (V_2-V_1)/L$, $f_2 := (V_1-V_0)/L$, $f_3 = (V_2-V_0)/L$. 
%
%
%
\begin{figure}[t]
\begin{center}
\scalebox{0.23}{\includegraphics{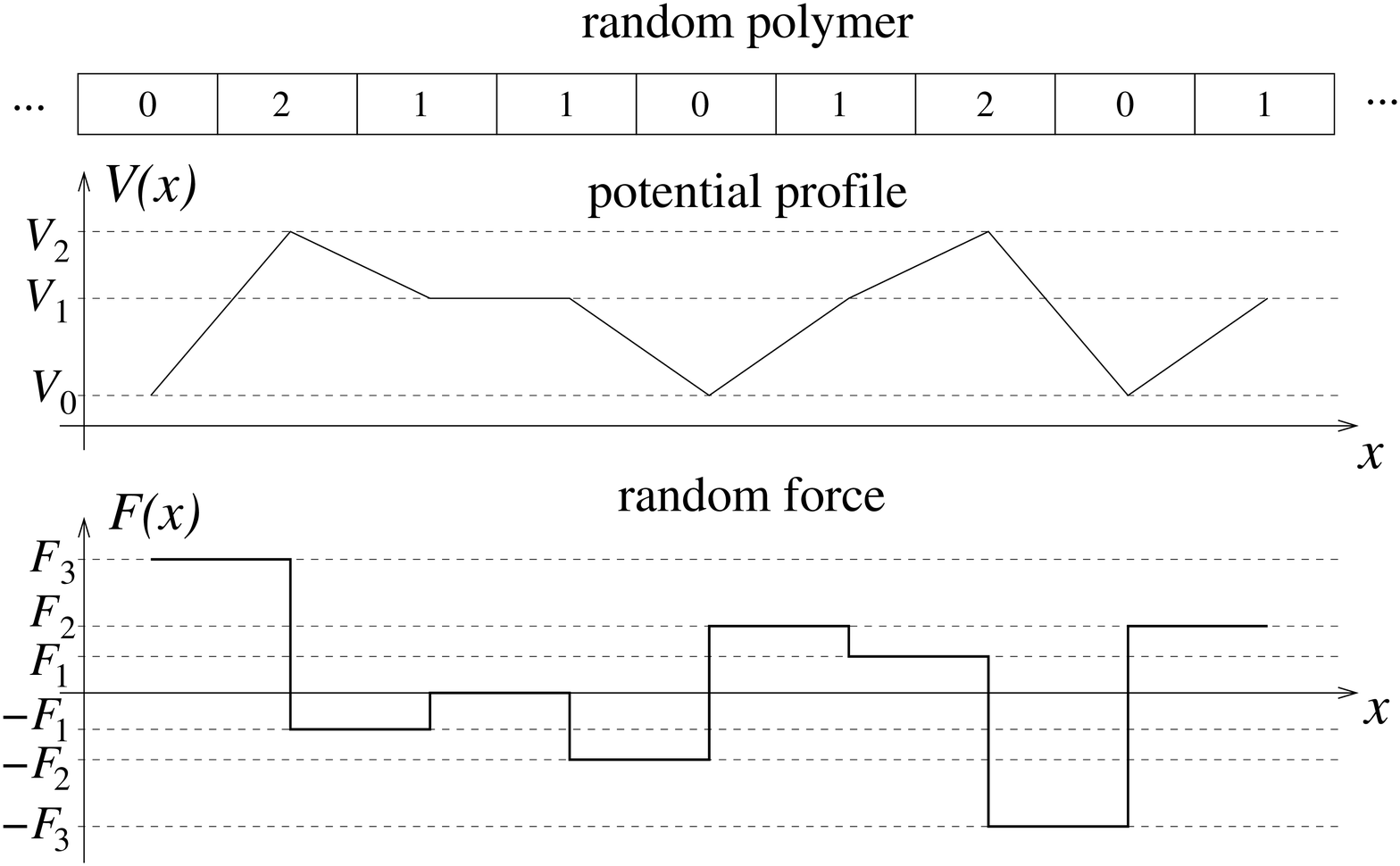}}
\end{center}
     \caption{
   Schematic representation of the piecewise linear model for the potential profile. We show the potential profile corresponding to a specific realization of the random polymer. We assumed that every monomer type has a given  ``potential value''. To go from one monomer to the adjacent one, the particle only has to pass the potential barrier which is the difference between the corresponding potential values associated to such monomers. It is also shown the corresponding gradient of the potential (the random force).      
   }
\label{fig:potential_profile}
\end{figure}
%
%
The above matrix means that, if the particle is going from a monomer $a$ to a monomer $b$, the force exerted over the particle in the meanwhile is the matrix element  $F_{a,b}$. A schematic representation of this situation is illustrated in Fig.~\ref{fig:potential_profile}. Notice that if the polymer consists of only one type of monomers, the gradient of the potential is zero everywhere because the  potential that feels the particle is constant along all the chain. 

If we put a particle on a realization of the random polymer and if we impose an external forcing $F_0$, we can write the function $\tau$ in terms of the matrix force $F$. Indeed, given  $a,b\in \{0,1,2\}$ we write,
\[
\tau_{a,b} := \frac{L}{F_{a,b}+F_0}.
\] 
Notice that, in order for $\tau$ to be finite we need that the external forcing be such that $F_0>\max_{a,b}\{|F_{a,b}|\}$. Now, to calculate the diffusion coefficient, we need first to calculate the expected value and the variance of $\tau$ as well as its corresponding correlation function.  Since our system is a Markov chain, such quantities can be written down straightforwardly~\cite{levin2009markov},
\begin{eqnarray}
\overline{\tau} &=& \sum_{a,b} \pi_a  \tau_{a,b} P_{a,b},
\label{eq:mean_tau_MC}
\\
\mbox{Var}[\tau]  &=& \left( \sum_{a,b}  \pi_a \tau_{a,b}\tau_{a,b}P_{a,b} \right) -\overline{\tau}^2,
\label{eq:var_tau_MC}
\\
C_\tau(\ell ) &=& 
\sum_{a,b}\sum_{a^\prime, b^\prime} \pi_a\tau_{a,b}\tau_{a^\prime,b^\prime} P_{a,b}P^{\ell-1}_{b,a^\prime}P_{a^\prime,b^\prime} -\overline{\tau}^2.
\label{eq:corr_tau_MC}
\end{eqnarray}
In terms of these quantities we can calculate $\varrho_\tau^2$ as 
\begin{equation}
\label{eq:varrho_tau_MC}
\varrho^2_{\tau} = \mbox{Var}[\tau] + 2 \sum_{\ell = 1}^\infty C_\tau (\ell).
\end{equation}
Since the correlations of a bounded observable vanishes exponentially fast in a Markov process, we have that $\varrho_\tau^2$ is always finite. 

\begin{figure}
\begin{center}
\scalebox{0.45}{\includegraphics{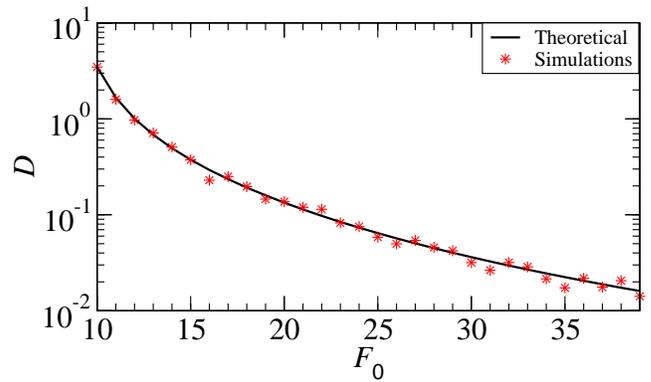}}
\end{center}
     \caption{(Color online)
     Diffusion coefficient for a particle placed on a random potential generated by a Markov chain. We plot (solid line) the diffusion coefficient as a function of the strength of the tilt $F_0$ obtained by the exact formula  Eq.~\eqref{eq:Deff}. This was obtained by numerically calculating the correlation function for $\tau$ given by Eq.~\eqref{eq:corr_tau_MC} and then substituting into the expression~\eqref{eq:varrho_tau_MC}. We also plot (red stars) the diffusion coefficient obtained directly from numerical simulations of the deterministic equation~\eqref{eq:overdamped_particle} for an ensemble of 750 particles during a time of $2000$ arb. units.  
    }
\label{fig:Deff_MarkovChain}
\end{figure}

For the simulations we adopt the values $V_0 = 1$, $V_1 = 3$, $V_2 = 10$ and $L = 1$. With these potential values we numerically calculate the matrix of crossing times for several values of the tilt $F_0$. Later we use this matrix to compute the mean value of $\tau$, its variance and the correlation function of that observable. This enables us to obtain the theoretical prediction for the diffusion coefficient, given by Eq.~\eqref{eq:Deff} through the sum of correlations for $\varrho_\tau^2$ (Eq.~\eqref{eq:varrho_tau_MC}), for the corresponding values of the tilt. Additionally we perform numerical experiments to determine the diffusion coefficients in order to compare with the respective theoretical predictions. This is shown in Fig.~\ref{fig:Deff_MarkovChain} which lets us appreciate a good agreement within the accuracy of our simulations.

%
%
%
%
%
%

\subsubsection{The expansion-modification system}

The E-M system is a stochastic process that was introduced as a simple model exhibiting spatial $1/f$ noise~\cite{li1989spatial,li1991expansion} and in particular was used to understand the long-range correlation present in DNA sequences~\cite{li1992long,li1994understanding}. Such a system is defined by two fundamental process as follows. Considering a ``seed'' (a symbol in the  binary alphabet $\{0,1\}$) it is  expanded with a probability $p$ and it is modified (it is changed to the other symbol) with the complementary probability $1-p$. In other words, if $x\in \{0,1\}$, this symbol is subjected to the process (extended coordinate-wise to words of arbitrary length),
\[
x \mapsto \left\{\begin{array}{cl} \overline{x} & \text{ with probability } p,\\ 
                                   x x & \text{ with probability } 1-p,
                  \end{array}\right.
\]
where $\overline{x}$ stands for the complementary symbol, i.e., $\overline{1} = 0$ and $\overline{0} = 1$.
This process allows the seed to grow in such a way that the resulting string of symbols becomes infinite with probability one.  This stochastic process has been studied within the context of symbolic dynamics in Ref.~\cite{salgado2012expansion}. In that work it was rigorously proved that the E-M system has a unique stationary measure having a polynomially decay of correlations for an open set of values of the parameter $p$. In the same work the authors conjectured that this behavior for the correlation functions occurs for almost all the values of the expansion probability and give an explicit formula for the  corresponding exponent $\beta(p)$ given by,
\begin{equation}
\label{eq:exponent_EM}
\beta (p) = \frac{ \log(1+p)-\log(|2p-1|)-\log(|3p -1|)}{ \log(1+p)}.
\end{equation}

Consider a particle on a polymer made up from two kinds of monomers '0' and '1' following the E-M process. For this case we will assume that the interaction potential particle-polymer depends on two monomers on the chain. We associate to a given dimer $ab$ a ``potential value'' as follows: given $a,b \in\{0,1\} $, let $\tilde V (a,b)$ be the potential associated to the dimer $ab$ given by
\begin{equation}
\label{eq:potential_EM}
\tilde V(a,b) = V_{a+b},
\end{equation}
where $V_i$ is a real number for every $i=0,1,2$. Notice that the index $i$ in the above definition is determined by the sum $a+b$ which can take only the values $0,1$ or $2$. This results in a potential profile which is piece-wise linear and depends on two-monomers. Thus, in this case the gradient of the potential depends on three symbols, and so does the corresponding crossing time $\tau$ which is given by
\[
\tau(a,b,c) : = \frac{L}{ \tilde V(a,b) - \tilde V(b,c) +F_0}.
\]
This equation can be interpreted in the following sense. The time that the particles spend traveling across the monomer $b$ is influenced by the presence of the adjacent monomers. The particle feels a potential due to its interaction with the triplet $abc$ while staying at $b$, where $a$ stands for the monomer behind $b$ and $c$ for the one in the front.
 
Since the correlations are polynomially decaying, we expect normal diffusion only when the corresponding exponent $\beta(p)$ is greater than $2$. According to Eq.~\eqref{eq:exponent_EM} we have that the expansion probability for which this occurs should be a value in the range $[0,p_1]$ where $p_1$ is approximately $p_1 \approx 0.833$.

\begin{figure}[t]
\begin{center}
\scalebox{0.38}{\includegraphics{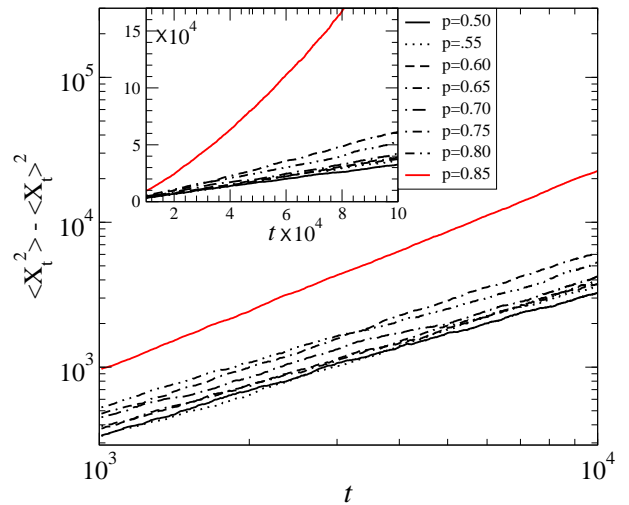}}
\end{center}
     \caption{(Color online)
     Mean square displacement as a function of $t$ for a disordered potential obtained from the E-M system.  We see that for the expansion probabilities $p=0.60,0.65,0.70,0.75,0.80$ (black lines) the diffusion is normal.  On the contrary, we can appreciate that for $p=0.85$ (red line) the normal diffusion does not longer holds. The latter is a consequence of the fact that the correlation exponent for such a expansion probability  is lower than one,  $\beta(0.85) < 1$. The latter implies that the CLT does not longer holds and consequently the diffusion coefficient diverge. 
    }
\label{fig:NormalDiffusion_EM}
\end{figure}

In Fig.~\ref{fig:NormalDiffusion_EM} we show the mean-square-displacement curves obtained by a simulation of an ensemble of $ 3000 $ particles obeying Eq.~\eqref{eq:overdamped_particle}. For these simulations we used the values for $V_0 = 1$, $V_1=3$, $V_2=10$ and $L=1$. We clearly see in this figure that the behavior of the mean square displacement is linear with $t$ or, in other words, there is normal diffusion. In the referred figure we plot the mentioned curves for the parameter values $p=0.60,0.65,0.70,0.75,0.80$ and $0.85$. We observe normal diffusion for the parameter values from $0.60$ to $0.80$. For $p=0.85$ the linear relation between the time and the mean square displacement it is no longer preserved, because, for this case, the CLT breaks down due to the slow decay correlations.  For that value of the expansion probability, we have $C_\tau(\ell) \propto \ell^{\beta(p)}$ where $\beta(0.85) < 1$ according to Eq.~\eqref{eq:exponent_EM}. Indeed it is clear that for such a decay of correlation the condition~\eqref{eq:summable_corr} no longer holds.

%
%
%
%
%
%

\section{Anomalous Diffusion}
\label{sec:AnomalousDiffusion}

In the preceding section we have shown that the normal diffusion occurs when the CLT holds. In order for the CLT to be valid for the ergodic sum of a given observable is sufficient to fulfill at least the two following conditions~\cite{chazottes2012fluctuations}: $i$) that the correlations in the system decay sufficiently fast and $ii$) that the observables are regular enough (e.g., bounded and  H\"older continuous). Here we will show that anomalous diffusion can arise when any of these conditions is broken.

\subsection{Slow decay of correlations}

We have pointed out that, in order for the CTL to hold for a regular observable it is sufficient that condition~\eqref{eq:summable_corr} be valid in order for the diffusion coefficient to be finite. In the case when this is not true we expect to have anomalous diffusion because $\varrho_\tau^2  = \infty$ which consequently implies that $D = \infty$. If this is the case we do not have a strong enough result analogous to the CLT unless for specific cases. For example, for the Maneville-Pomeau map, it is known to have polynomial decay of correlations with a given exponent ranging from zero to infinity. Moreover, it has been proved that, generically a given observable will satisfy a CLT for correlation exponent $\beta > 1$. On the contrary, if the correlation exponent $\beta$ is in the interval $(0,1]$ a stable limit law holds for regular observables~\cite{chazottes2012fluctuations,gouezel2004central}. This result was stated in the general framework of abstract Markov maps~\cite{gouezel2004central} having some ergodic and mixing properties. In the case of stochastic process there is no analogous results known by the authors. 

In the example of the preceding section we obtained disordered potentials  by means of the E-M process. For  this system it has been proved rigorously to have long range decay of correlations, a property which makes it similar (in the stochastic sense) to the Maneville-Pomeau map. Using this analogy we can hypothesize that for the E-M system, the ergodic sums (appropriately normalized) might converge to a stable law.  This would happen when the correlations decay  as $\ell^{-\beta}$ with $ 0 < \beta <1$. Using the above hypothesis the total time $T_n$ that the particle spends to cross throughout $n$ monomers of the chain, will converge to a stable law. Equivalently, we have that,
\[
\frac{T_n-n\overline{\tau}}{ n^\alpha} \to W, \quad \mbox{ as } \quad  n\to \infty,
\]
where $W$ is a random variable with a stable law. Here $\alpha$ is the exponent of the normalization sequence $\{n^\alpha\}_{n=0}^\infty$ which, according to the case of the Maneville-Pomeau map, depends on the correlation exponent as
\begin{equation}
\label{eq:exp_alpha}
\alpha(p) = \left\{
 \begin{array}{ll}
  \frac{1}{1+\beta(p)} & \quad \text{if } \beta(p) < 1,\\
   1/2  & \quad  \text{if }\beta(p)>1.
 \end{array} \right.
\end{equation}

As in the case of the normal diffusion, we can use the above hypothesis to estimate the asymptotic behavior of the total displacement of the particle during a time $t$. If we denote by $N_t$  the random variable defined as the number of monomers that the particle has crossed during a time $t$, we prove in Appendix~\ref{ape:Dist_N} that,
\[
N_t \to \frac{t}{\overline{\tau}} + \frac{ t^\alpha }{\overline{\tau}^{1+\alpha}} W,
\]
as $t\to\infty$. From this is result, it is clear that the square displacement of the particles goes as
\begin{equation}
\label{eq:MSD_EM}
(X_t - \mathbb{E}(X_t))^2 =  L^2\left(N_t - \frac{t}{\overline{\tau}}\right)^2 \approx \frac{ t^{2\alpha} }{\overline{\tau}^{2+2\alpha}} W^2,
\end{equation}
which means that the square displacement goes typically as $t^{2\alpha}$ implying anomalous diffusion for $\alpha>1/2$.

%
%
\begin{figure}[ht]
\begin{center}
\scalebox{0.36}{\includegraphics{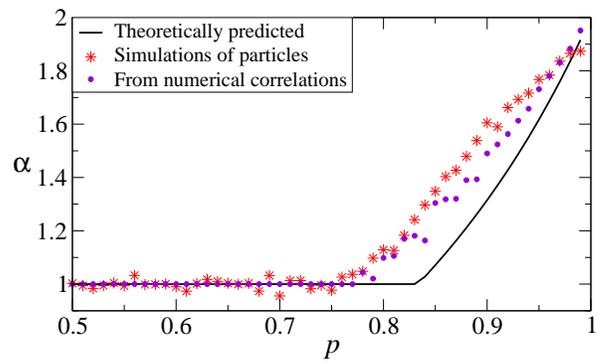}}
\end{center}
     \caption{(Color online)
     Transition from normal to anomalous diffusion in the E-M model. We plot the behavior of the diffusion exponent $2\alpha$ obtained from numerical simulations (red stars) of an ensemble of $4800$ particles obeying Eq.~\eqref{eq:overdamped_particle}. We also plot the theoretical prediction for the exponent given by Eq.~\eqref{eq:exp_alpha} (solid curve). The behavior of the exponent exhibits clearly the transition from normal to anomalous diffusion. The discrepancy between the theoretically expected exponents and those found by simulations might be due to finite size effects, as it was pointed out in Ref.~\cite{salgado2012expansion}. The numerically calculated correlation exponents from the potentials give diffusion exponents (violet filled circles) which agree with those obtained from direct simulations. 
    }
\label{fig:AnomalousDiff_EM}
\end{figure}
%
%

In order to test these results we performed numerical simulations. First,  given  a value of the expansion probability $p$ we  obtained several realizations of binary sequences. Next according to them we obtained the corresponding  potentials from Eq.~\eqref{eq:potential_EM}. Then we simulated the deterministic dynamics given by Eq.~\eqref{eq:overdamped_particle} of an ensemble of $4800$ particles during a time of $10^4$ arb.\ units. We calculated the average (over the ensemble) of the square displacement as a function of $t$. Finally, we performed a fit to a power law of the obtained curve to determine the corresponding exponent. This procedure was done for several values of the expansion probability and we plot the exponent as a function of this parameter in Fig.~\ref{fig:AnomalousDiff_EM}. In the same figure we also plot the theoretical prediction given by Eqs.~\eqref{eq:MSD_EM} and~\eqref{eq:exp_alpha}.  The  discrepancy observed in Fig.~\ref{fig:AnomalousDiff_EM} between the numerical results and the theoretical prediction can be understood as a manifestation of finite size effect. In Ref.~\cite{salgado2012expansion} is was pointed out that numerical simulations of the E-M system could lead to misleading observations because of large statistical errors. The latter is consequence of the very slowly convergence of the process to its stationary state. In our case we obtained the disordered potentials by simulating the E-M process up to a finite number of iterations. Then, the  corresponding correlation exponent present in the realization of the chain would no correspond to $\beta(p)$ given by Eq.~\eqref{eq:exponent_EM}. To see that this is the case in our simulations we calculated numerically the correlation exponent for the obtained sequences. Then we calculated the corresponding exponent of the average square displacement by means of Eq.~\eqref{eq:exp_alpha}. The result is also plotted in Fig.~\ref{fig:AnomalousDiff_EM}. We observe that the discrepancy from such exponents from the theoretical ones are of the same order of the discrepancy from the obtained by direct simulation of the deterministic dynamics. In this way we can say that our theoretical results and the hypothesis about the validity of the convergence to a stable law are consistent with the simulations.

\subsection{Unbounded crossing times with fast decay of correlations}
\label{subsec:Unbounded}

The preceding subsection have dealt with the case in which the CLT is broken down by the non-summability of the correlations. This is however, not the only way in which it is lost the validity of CLT. Another way to do that is by considering a unbounded observable. 
The observable to which the CLT is applied is the crossing time function $\tau$. If such an observable is no longer finite for some monomers, it is expected anomalous diffusion to occur. This is because the distribution of crossing time can have in a heavy tailed distribution and the CLT could be broken down~\cite{chazottes2012fluctuations} for a sufficiently small exponent. If we have no correlations we can apply a  classical result in probability theory about the sum of random variables belonging  to a domain of attraction which leads to a stable law~\cite{gnedenko1968limit}. This lets us characterize the exponent with which the mean square displacement diverges with time. Clearly this situation can occur in the case in which the external forcing $F_0$ is exactly the critical tilt, as it was pointed out in Ref.~\cite{kunz2003mechanical,denisov2010biased}. 

To illustrate this situation consider a system with a finite or countable infinite monomer types. Assume that a random potential, generated by some stationary process, is piece-wise linear and that the possible slopes are in a (finite or infinite) set $\{f_1, f_2, \cdots \}$. The critical tilt will be given by $F_c = -\inf\{f_1,f_2,\cdots\}$. If the polymer is such that the monomer realizing the supremum occurs with a positive probability (no matter how small), the particles in such a tilted potential will get stuck with probability one. This results in a system with zero diffusion coefficient since all the particles are eventually stopped. Observe that this always happens when we have a finite number of monomer types. Otherwise it is  necessary to consider a system with an infinite number of monomer types in which the supremum over all $f_i $'s has zero probability to occur. 

The system that we will consider here is the following. Assume that the slopes along the chain are realizations of independent and identically distributed random variables $\{K_m\}_{m=-\infty}^\infty$,  with state space $\mathcal{F} = \{\cdots, f_{-2}, f_{-1}, f_0, f_1, f_2, \cdots \}$. Assume additionally that $f_j=-f_{-j}$, and that the sequence $0 < f_1 <f_2<f_3 \cdots$ has a finite limit,
\[
f_\infty := \lim_{j\to \infty} f_j <\infty.
\]
Thus, the critical tilt is $F_c = f_\infty$. Let $K$ be a random variable with the same distribution of $K_m$ for all $m\in \mathbb{Z}$.  The probability function for  $K$ is assumed to be
 \begin{equation}
 \label{eq:prob_iid}
p(j) = \mathbb{P}(K = f_j) = \frac{2^{-|j|}}{3}.
\end{equation}
Next,  the time that the particle spends crossing from $m$-th monomer to the $(m+1)$-th monomer in the chain is a random variable denoted by $\Theta_m$. The force that such a particle feels in the meanwhile is $K_m$ and therefore,
\[
\Theta_m = \frac{L}{K_m + F_c}.
\]   
These random variables have state space given by $\mathcal{T} = \{\cdots, \tau_{-2}, \tau_{-1},\tau_{0},\tau_{1},\tau_{2},\}$, where,
\[
\tau_j := \frac{L}{f_j + F_c}.
\]
Notice that the set of random variables  $\{\Theta_m\}_{m=-\infty}^\infty$  are independent and identically distributed as the $K_m$'s. Let $\Theta$ be a random variable with the same distribution of $\Theta_m$ for all $m\in \mathbb{Z}$. Consider the quantity $\mathbb{P}(\Theta > t)$. It is easy to see that,
\begin{eqnarray}
\nonumber
\mathbb{P}(\Theta > t) &=& \mathbb{P}\left( \frac{L}{K + F_c} > t \right)
\\
\nonumber
&=& \mathbb{P} \left( K < \frac{L}{t} - F_c \right) = \sum_{j=-\infty}^{\kappa(t)} p(j),
\end{eqnarray}
where $\kappa(t)$ is an integer defined as
\[
\kappa(t) := \max\left\{ j \in\mathbb{Z} \, :  \, f_j < \frac{L}{t} - F_c  \right\}.
\]
Notice that if $t$ is large enough we have that $\frac{L}{t} - F_c$ is negative, and consequently $\kappa$ will be negative and thus,
\begin{eqnarray}
\mathbb{P}(\Theta > t) &=& \sum_{j=-\infty}^{\kappa(t)} p(j) = \sum_{j=-\infty}^{\kappa(t)} \frac{2^{-|j|}}{3},
\nonumber 
\\
&=& \frac{2^{-|\kappa(t)|}}{3} \sum_{m=-\infty}^{0} 2^{-|m|},
\nonumber
\end{eqnarray}
or equivalently,
\[
\mathbb{P}(\Theta > t) =  \frac{2 }{3}\, 2^{-|\kappa(t)|}.
\]

In order to have an explicit expression for $\kappa(t)$ consider the following model for $f_j$, 
\begin{equation}
\label{eq:fjs}
f_j = \mbox{sign}(j) f_\infty \left( 1 - 2^{-|j|/q}\right).
\end{equation}
where $f_\infty$ is a positive constant and $q>0$ is a parameter. A few calculations show that 
\[
2^{-|\kappa(t)|} = \left(\frac{L}{f_\infty}\right)^{q} \, t^{-q},
\]
which gives for the distribution of $\Theta$,
\begin{equation}
\label{eq:domainIII}
\mathbb{P}(\Theta > t) = \frac{2}{3} \left(\frac{L}{f_\infty}\right)^{q} \, t^{-q}.
\end{equation}
According to~\cite{gnedenko1968limit}, if a  random variable satisfies~\eqref{eq:domainIII} for $1<q<2$, then a sum of identically distributed random variables with the same distribution of $\Theta$ converges to a stable law. This result implies that the time that the particle spends crossing throughout $n$ monomers,
\[
T_n = \sum_{j=0}^{n-1} \Theta_j,
\]
is such that
\[
\frac{T_n - n\overline{\tau}}{n^{1/q}} \to W,
\]
where $W$ is a random variable with a stable law. In this case, the mean value $\overline{\tau} := \mathbb{E}[\Theta] $ is finite  for all $q>1$ as shown in Appendix~\ref{ape:bound_tau}. 

According to Appendix~\ref{ape:Dist_N} we  have that the convergence to a stable law implies anomalous diffusion with exponent $2/q$ for $1<q<2$.  It is clear that in our system the parameter $q$ is not restricted to  that values but it can be greater that $2$. Indeed for parameter values above $q=2$ we have that for $T_n$ the CLT holds~\cite{gnedenko1968limit}. Thus in this case we have normal diffusion and therefore a transition from anomalous to normal diffusion occurs when $q$ increases. 

\section{Discussion and Conclusions}
\label{sec:Conclusions}

We have studied the diffusion of particles moving deterministically on disordered correlated potentials. We found that for potentials with summable correlations occur normal diffusion as a consequence of the CLT. This lets us calculate the diffusion coefficient  in terms of the sum of correlations of crossing times, which is given by Eq.~\eqref{eq:Deff}.  To test the last expression, we produced random polymers by means of a stationary Markov process with three kinds of monomers. On this polymer we simulated an ensemble of deterministic particles and then we calculated the diffusion coefficient.  We found normal diffusion which agreed with our theoretical results within the accuracy of our numerical simulations. We also performed the same simulations with a polymer produced in different way. For this purpose we used the E-M system to obtain binary sequences with long-range correlations of polynomial type. In this case we also found normal diffusion when the correlation  decays faster than $\ell^{-1}$. This was consistent with the fact that the correlation must be summable in order for the CLT to be valid.

Next we gave two examples where anomalous diffusion is expected to occur when the CLT is no longer valid. Indeed for a system in which the speed of decay of correlations is controlled  by a parameter, we expected a transition from normal to anomalous diffusion if the correlation exponent can go down (continuously) to zero. This occurs for the classical Maneville-Pomeau map which is the only system (up to our knowledge) for which rigorous results are available. Here we present a different example (E-M system) for which this control is possible.

The first case we explored was the presence of sufficiently slow decay of correlations. For such a purpose we used the E-M system to generate random potentials with correlation decaying slower than $\ell^{-1}$. Next, by assuming that the total time to cross throughout $n$ monomers converge to a stable law for large $n$ (as in the case of the well-known Maneville-Pomeau map) we obtained an analytical expression for the diffusion exponent. Then we performed numerical simulations of the deterministic dynamics finding a transition from normal to anomalous diffusion, which agreed with our theoretical results. Finally we gave an example of a system with unbounded crossing times where we also obtained anomalous diffusion. This was the case in which the random potentials have infinitely many slopes and the external force equals the critical value. 

In conclusion  we provided a fairly general framework to deal with the problem of deterministic overdamped transport in disordered potentials. We gave a formula for the diffusion coefficient when normal diffusion occurs. Condition for the latter were stated over the correlations of the potentials. We also gave examples where transitions from normal to anomalous diffusion happens. In the last case we assumed that ergodic sums converge to a stable law.  All these results give us plausible arguments to say that this transition might occur due to a breakdown of the CLT.

\section*{Acknowledgements} 

RSG thanks CONACyT by financial support through Grant No. CB-2012-01-183358. CM thanks PROMEP for financial support by the  scholarship UASLP-CA-188. The authors thank E. Ugalde for useful discussions.

\appendix 

\section{Position distribution of constant traveling time}
 \label{ape:Dist_N}

In this section we prove that the random variable $N$ defined as the number of cell that the particle has traveled during a time $t$ has a normal distribution with mean $t/\overline{\tau}$ and variance $Dt$. Such a random variable is defined implicitly by the equation $T_{N_t} = t $ or equivalently by,
\[
T_{N_t} = \sum_{j=0}^{N_t} \tau(\sigma^j (\mathbf{a})) = t.
\]
If the ergodic sum $T_n$ converge to a stable law or to a normal distribution we have that for large enough $n$, 
\begin{equation}
\label{eq:stable-law-Tn}
\frac{T_n - n\overline{\tau}}{n^\alpha} \to W, \quad \mbox{ when } \quad n \to \infty,
\end{equation}
where $W$ is a random variable with a stable law (for which the variance is infinite) if $\alpha > 1/2$ or a random variable with a normal distribution (with finite variance $\varrho_\tau^2$ given by Eq.~\eqref{eq:var_tau}) if $\alpha = 1/2$. The value of $\alpha$ depends on the properties of the function $\tau$. It is clear that the statistical properties of $T_{n}$ defines the statistical properties of $N_t$. The CLT gives an ``approximate'' distribution for $T_n$ through Eq.~\eqref{eq:stable-law-Tn} and from this approximation it can be inferred the distribution for $N_t$. If take the equality in Eq.~\eqref{eq:stable-law-Tn} and put $T_n = t$ and $n = N_t$, we have that,
\begin{equation}
\label{eq:implicit_N_t}
\frac{t - N_t\overline{\tau}}{N_t^\alpha} = W.
\end{equation}

Notice that the random variable $W$ is a distribution centered at zero, i.e., the most probable values of $W$ are around zero. This means that the random variable $N_t$ will have a distribution centered around the  root of the function $ \psi (N_t) := (t - N_t\overline{\tau})/N_t^\alpha$ which is given by $N_t = t/\overline{\tau}$. Since $W$ is fixed, we can solve for $N_t$ in terms of $W$ using a linear expansion of $\psi(N_t)$ around $N_t = t/\overline{\tau}$. We obtain
\[
\psi (N_t) =  -\frac{\overline{\tau}^{1+\alpha}}{t^\alpha} \bigg(N_t - \frac{t}{\overline{\tau} }\bigg) + \mathcal{O} (t^{-\alpha-1}), 
\]
which implies that,
\[
 -\frac{\overline{\tau}^{1+\alpha}}{t^\alpha} \bigg(N_t - \frac{t}{\overline{\tau} }\bigg) \approx W,
\]
for $t$ large enough. From the above we can observe that, 
\[
N_t \approx \frac{t}{\overline{\tau}} -\frac{t^\alpha}{\overline{\tau}^{1+\alpha}} \, W.
\]
Notice that the expected value of $N_t$ is $t/\overline{\tau}$ since $\mathbb{E}[W] =0$. Then we can see that the square deviation of $N_t$ around its mean value $t/\overline{\tau}$ has typical values that goes as $t^{-\alpha}$, i.e., 
\begin{equation}
\left(N_t - \frac{t}{\overline{\tau}}\right)^2 \approx \frac{t^{2\alpha}}{\overline{\tau}^{2+2\alpha}} \, W^2.
\end{equation}

In the case $\alpha = 1/2$ we have that $W$ has finite variance equals $\varrho_\tau^2 $. This means that the variance of $N_t$ is finite (the mean square displacement is well defined) and is given by,
\begin{equation}
\mbox{Var}(N_t) = \mathbb{E}\left[ \left(N_t - \frac{t}{\overline{\tau}}\right)^2 \right]  \approx \frac{\varrho_\tau^2}{\overline{\tau}^{3}} \, t,
\end{equation}
 from which it follows straightforwardly that the diffusion coefficient is given by Eq.~\eqref{eq:Deff}.

\section{Upper bound of $\overline{\tau}$ for unbounded crossing times. }
 \label{ape:bound_tau}

For the unbounded crossing time $\tau$ presented in Section~\ref{subsec:Unbounded} we have that its expected value is given by,
\[
\overline{\tau} = \sum_{j=-\infty}^\infty \tau_j \, \mathbb{P}(\Theta = \tau_j)  = \sum_{j=-\infty}^\infty  \tau_j(f_j ) \, \mathbb{P}(K = f_j),
\]
which is equivalent to
\[
\overline{\tau}  = \sum_{j=-\infty}^\infty  \frac{L}{f_j + F_c} \, \frac{2^{-|j|}}{3}.
\]

The above sum can be split into two sums
\begin{eqnarray}
\nonumber
\overline{\tau} &=&   \sum_{j=-\infty}^0  \frac{L}{f_j + F_c} \, \frac{2^{-|j|}}{3} + \sum_{j=1}^\infty  \frac{L}{f_j + F_c} \, \frac{2^{-|j|}}{3}.
\end{eqnarray}
Call $I_-$ and $I_+$ the first and the second sums, respectively, in the above equation. Notice that $I_-$ can be done exactly. Substituting the explicit form of $f_j$ given in Eq.~\eqref{eq:fjs} in $I_-$ yields,
\begin{eqnarray}
\nonumber
I_- &=&  \frac{2L}{3F_c} \sum_{j=-\infty}^0 \frac{ 2^{-|j|} }{ 2^{ -|j|/q} }  
\\
\nonumber
&=&\frac{2L}{3F_c}  \frac{ 1 }{ 1 - 2^{ - 1 + 1/q} },  
\end{eqnarray}
which is clearly bounded for all $q>1$. 

Substituting the explicit form of $f_j$ given in Eq.~\eqref{eq:fjs} in the second sum $I_+$ we observe that,
\begin{eqnarray}
\nonumber
I_+ &=&   \frac{L}{3F_c} \sum_{j=1}^\infty \frac{ 2^{-|j|} }{1 - 2^{ -|j|/q-1} }  
\\
\nonumber
&\leq&  \frac{L}{3F_c} \sum_{j=1}^\infty \frac{ 2^{-|j|} }{1 - 2^{ -1} }  = \frac{2L}{3F_c},
\end{eqnarray}
this shows that $\overline{\tau}$ is bounded from above for all $q>1$.

\nocite{*}

\bibliography{NormalAnomalousRefs}

\end{document}